\PassOptionsToPackage{dvipsnames}{xcolor}

\documentclass[manuscript]{acmart}

\usepackage{hyperref}
\usepackage{xcolor}
\usepackage{multirow}
\usepackage{wrapfig}

\setcopyright{none}
\AtBeginDocument{%
  \providecommand\BibTeX{{%
    \normalfont B\kern-0.5em{\scshape i\kern-0.25em b}\kern-0.8em\TeX}}}

\copyrightyear{2024} 
\acmYear{2024} 
\setcopyright{rightsretained} 
\acmConference[LAK '24]{The 14th Learning Analytics and Knowledge Conference}{March 18--22, 2024}{Kyoto, Japan}
\acmBooktitle{The 14th Learning Analytics and Knowledge Conference (LAK '24), March 18--22, 2024, Kyoto, Japan}
\acmDOI{10.1145/3636555.3636880}
\acmISBN{979-8-4007-1618-8/24/03}

\acmSubmissionID{178}

\begin{document}

\title{The Sequence Matters in Learning - A Systematic Literature Review}

\author{Manuel Valle Torre}
\email{m.valletorre@tudelft.nl}
\orcid{0000-0002-0456-8360}
\affiliation{%
  \institution{Delft Institute of Technology}
  \streetaddress{Van Mourik Broekmanweg 6}
  \city{Delft}
  \country{Netherlands}
  \postcode{2628 XE}
}
\author{Catharine Oertel}
\email{c.r.m.m.oertel@tudelft.nl}
\orcid{0000-0002-8273-0132}
\affiliation{%
  \institution{Delft Institute of Technology}
  \streetaddress{Van Mourik Broekmanweg 6}
  \city{Delft}
  \country{Netherlands}
  \postcode{2628 XE}
}
\author{Marcus Specht}
\email{m.m.specht@tudelft.nl}
\orcid{0000-0002-6086-8480}
\affiliation{%
  \institution{Delft Institute of Technology}
  \streetaddress{Van Mourik Broekmanweg 6}
  \city{Delft}
  \country{Netherlands}
  \postcode{2628 XE}
}

\renewcommand{\shortauthors}{Valle Torre, et al.}

\begin{abstract}
Describing and analysing learner behaviour using sequential data and analysis is becoming more and more popular in Learning Analytics.
Nevertheless, we found a variety of definitions of learning sequences, as well as choices regarding data aggregation and the methods implemented for analysis. 
Furthermore, sequences are used to study different educational settings and serve as a base for various interventions. 
In this literature review, the authors aim to generate an overview of these aspects to describe the current state of using sequence analysis in educational support and learning analytics. 
The 74 included articles were selected based on the criteria that they conduct empirical research on an educational environment using sequences of learning actions as the main focus of their analysis.
The results enable us to highlight different learning tasks where sequences are analysed, identify data mapping strategies for different types of sequence actions, differentiate techniques based on purpose and scope, and identify educational interventions based on the outcomes of sequence analysis.
\end{abstract}




\maketitle

This version is for personal use and not for redistribution.
The final version was published as part of the proceedings of the 14th Learning Analytics and Knowledge Conference (LAK '24).
March 18--22, 2024, Kyoto, Japan.
\href{https://doi.org/10.1145/3636555.3636880}{https://doi.org/10.1145/3636555.3636880}

\section{Introduction}
Educational Technologies (EdTech) have rapidly scaled teachers' capacity for instruction, for example through live, virtual and recorded lectures \cite{elaouifi2021PredictingLearnerPerformance, mubarak2021PredictiveLearningAnalytics}.
EdTech,  has also increased the opportunity for learners to actively engage with learning materials, from practice problems in electronic environments to automated assessments, or annotation tools \cite{dever2022PedagogicalAgentSupport, li2022TemporalStructuresSequential}.
However, some processes that link teaching and learning in a continuous cycle, such as formative assessment, and personalized support and feedback, have not scaled at the same rate \cite{sedrakyan2020LinkingLearningBehavior}. 
From its origins, Learning Analytics (LA) intends to close this cycle using data generated by learners in educational systems, and processing it to obtain insights grounded in learning theory and provide (semi) automated learning support and feedback \cite{clow2012LearningAnalyticsCycle}.

Existing implementations of LA, such as dashboards or nudges, are often based on descriptive analytics of performance metrics \cite{valletorre2020EdXLogData}, designed for summative assessment and (self) monitoring \cite{wong2019ExploringSequencesLearner}. 
However, analytics of the learning process with the detail level to enable timely assessment and support are an ongoing development \cite{sedrakyan2020LinkingLearningBehavior}.
Learning is a process that unfolds and changes over time, and aggregating or counting the actions the learner executes is limited in its narrative power \cite{saint2020CombiningAnalyticMethods}.
For that reason, it is critical to understand the execution of learning tasks as a temporal and sequential notion \cite{molenaar_temporal_2022}.
Using sequential data to observe, understand, and support learning is not exclusive to LA, but has a history in fields such as Educational Data Mining (EDM), Intelligent Tutoring Systems (ITS), and Artificial Intelligence in Education (AIED) \cite{wang2023SystematicReviewEmpirical}.
This results in a lack of common definitions of concepts such as learning sequences or sequence units.
Furthermore, there are no established best practices of the analysis methods that can be applied for specific purposes, educational settings or scenarios \cite{knight2017TimeChangeWhy}.
As a consequence, research is often in the exploratory stage, using data from scenarios where the analysed course or task is over, without intervention \cite{motz2023LAKDirectionMisalignment}.

In this article, we work towards a broader understanding of the variety of definitions, data points, methods, and interpretations used for the analysis of learning as a sequence. 
We aim to enable a broader understanding of the learner context to be taken into account for intelligent feedback and analytics interventions. 
In the following, we will first introduce the state-of-the-art of learning sequence analytics and then present the analysis of 74 articles considering the given dimensions.

\section{Sequence Analysis and Learning Sequences}
Sequence Analysis (SA) of human behaviour, where a sequence is an ordered list of events, allows the contextualization of these events, their relationships with each other, and the effects of part of or the whole sequence \cite{abbott1995SequenceAnalysisNew}.
SA is especially useful to research phenomena that occur as a process across time, where the use of static attributes and frequency rates may not be informative enough to understand their development \cite{abbott1995SequenceAnalysisNew}.
The analysed sequences can include actions executed by a person as well as reactions of others in their environment, i.e. a peer or an intelligent agent.

Sequences can have different interpretations in education, in this review, we focus only on the sequences of actions that learners execute during a learning task.
For example, the steps followed to solve a problem \cite{wang2023SystematicReviewEmpirical} or the sequences of answers that learners provide in guided practice environments, and how certain sequences may relate to learners' behaviours \cite{venant2017UsingSequentialPattern} or characteristics \cite{baker2010DetectingGamingSystem}.

\subsection{Sequences of Actions in Learning Tasks}
Learning tasks refer to situations when learners actively engage with the learning materials which require them to integrate domain-specific knowledge and procedures with more general critical thinking and analysis skills \cite{nokes2010ProblemSolvingHuman}, such as solving complex problems \cite{kirschner2008TenStepsComplex} or scientific inquiry tasks \cite{wang2023SystematicReviewEmpirical}. 
For tasks in different domains, there are behavioural patterns that appear with different levels of practice.
For example, experts in a domain may show deeper categorization of a problem, use forward-thinking strategies in their execution of the solution or iteratively evaluate intermediate solutions \cite{nokes2010ProblemSolvingHuman}.
In these tasks, learners have a large number of operators or actions available in the problem space, or learning environment, making the combinations leading to a solution difficult to observe.
Furthermore, defining and describing such combinations is not trivial, since they are "automated", or stored in one's memory as procedural learning, making it difficult to explain by experienced individuals \cite{nokes2010ProblemSolvingHuman}.
As a consequence, the assessment and monitoring of learners are particularly difficult for teachers, especially considering the scale provided by educational technologies and learning platforms \cite{hurtig2022NetworkVisualizationAssessment}.
The data produced by these systems, however, can be leveraged to scale the capabilities of teachers, by analysing sequences of learner's data records to identify behaviours and reveal learning processes \cite{matcha2019DetectionLearningStrategies}. 

\subsection{Trace Data Sequences and Analysis}
The reliability of trace data comes from its ability to express dynamic events, more than as aggregated, static aspects of a measure \cite{wang2023SystematicReviewEmpirical}.
Using analytics to find patterns in the sequences that indicate specific tactics or common issues can help practitioners obtain insights into the use of different strategies.
Such patterns can be a set of learning actions that occur in succession for a proportion of the learners, from partial segments to complete sequences.
In addition, given that the patterns are extracted from sequences of actions, they can be interpreted by the teacher and even learners themselves \cite{tang2021ModelingLearningBehaviors}. 
For example, in a 3D geometry game where learners can insert, remove, and rotate shapes to solve puzzles, teachers cannot assess nor give feedback to learners based on the gameplay records without assistance.
Common patterns that lead to a failed assignment can be identified as mistakes and teachers can prioritize which to address \cite{gomez2021ApplyingLearningAnalytics}.

Analysing sequences of trace data throughout a learning task offers practical advantages as well \cite{taub2019HowDoesPrior}: when systems process data traces as sequences of learner actions, there is a constant input of information to incrementally update the learner model, enabling the system with the flexibility to constantly adapt to the learner's level or intervene whenever necessary \cite{corbeil2020ProbabilisticApproachesDetect}.
This allows practitioners, and eventually, an intelligent system, to predict if the current path is trending towards an undesirable outcome, to determine which action patterns are leading to that result and how to leverage these insights to optimise learning \cite{tang2021ExploratoryAnalysisLatent, riofrio-luzcando2016StudentActionPrediction}.

\subsection{Related Work}
\label{sec:related}
The analysis of learning sequences is emerging from several fields, and while there are no surveys specifically on the topic, there are related works from the EDM, LA and AIED communities. 
Regarding the use of educational data to observe learner behaviour and predict performance, Xiao et al. \cite{xiao2022SurveyEducationalData} analyse EDM methods in the age of big data to discover the trends around the complete data mining pipeline, such as data collection and preprocessing, feature engineering, classifier and ensemble method selection, and interpretation of results.  
While they mention techniques that analyse sequential data, there are no definitions of learning sequences or sequence units.
On the other hand, the survey by Bogarín et al. \cite{bogarin2018SurveyEducationalProcess} focuses on Educational Process Mining (EPM), a field that studies the process-oriented view of EDM, particularly as an end-to-end process.
The authors describe Virtual Learning Environments (VLE) and the Event Logs they produce as valuable data sources for sequential analysis, as they are detailed records of learners' actions.
They mention methods handling learning as sequences, including Sequential Pattern Mining, but the definitions outside of process mining are not addressed.

Regarding the use of trace data to observe learners' actions, the review by Wang et al. \cite{wang2023SystematicReviewEmpirical} analyses research on the use of log data from Open-Ended Learning Environments (OELEs) to understand scientific inquiry and problem-solving. 
Their conceptual framework consists of 4 points: the competencies measured, mainly problem-solving and inquiry; the observations used, ranging from multiple-choice questions to log data; the feature extraction and modelling processes and the interpretation of the analysis.
While the focus is not only on sequential methods, almost half of the 70 reviewed papers use sequential features for analysis, with 29 applying pattern mining to identify behaviours.
The review by Nadimpally et al. \cite{nadimpalli2023SystematicLiteratureReview} surveys Artificial Intelligence techniques for adaptive learning systems, identifying two main types: to model learner behaviour and to organize learning content.
Concerning modelling learner behaviour, they found that the most common methods are Similarity-based Pattern Recognition, Probabilistic Graphical Models, and Deep Learning: all of them capable of processing sequential data.
Existing literature explores the growing use of temporal data to identify, classify or predict learner behaviours, which often involves sequence analysis.
However, a focused study of when to use sequence analysis, the data units that can be leveraged and which methods to implement, is less covered. 

\section{Research Questions}
\label{sec:objectives}
Addressing the shortcomings from the previous section, this study intends to establish a common ground for Sequence Learning Analytics from the literature, so researchers can use this to find similar works, compare their approach to the state-of-the-art, and leverage insights of existing work, allowing the community to also focus on the intervention design.
In this review, a \textbf{learning sequence} is defined as an ordered list of context-specific actions, with each \textbf{unit} being a systematically mapped learning action from one or more data traces.
The following questions guide our analysis: 
\begin{enumerate}
    \item How are learning sequences analyzed to describe and support learners' actions in computer systems? 
    \begin{enumerate}
        \item Which types of \textit{learning tasks} are analysed as sequences? 
        \item What is the motivation or \textit{purpose} when using learning sequences? 
        \item Which \textit{actions} are used as the learning sequence units?
        \item Which \textit{methods} are used to analyse learning sequences? And what are the interpretations of the patterns that result from them?
        \item Which \textit{educational interventions} are designed with the obtained insights? Which learning principles are used to support and evaluate them?
    \end{enumerate}
\end{enumerate}

\section{Method}
\label{sec:methods}
In this review, we followed the Preferred Reporting Items for Systematic reviews and Meta-Analyses (PRISMA) \cite{page2021PRISMA2020Statement}, to answer the questions above, in Section \ref{sec:objectives}. 
The queries, information sources and eligibility criteria are described in this section, followed by the document collection and inclusion process, and the data extraction method.

The queries were executed in SCOPUS and Web of Science (WoS), for the domains of \textcolor{Cerulean}{Educational Data Mining (EDM), Artificial Intelligence in Education (AIED), Learning Analytics (LA), Intelligent Tutors and Learner Modeling}, exploring the fields of title, abstract and author-defined keywords. 
SCOPUS and WoS were selected as they include most of the top publications in such domains, such as the proceedings of the conferences on Learning Analytics and Knowledge, Educational Data Mining, and the International Conference on Computers in Education, as well as the Journal of Learning Analytics, British Journal of Educational Technology, Journal of Educational Data Mining, and Computers and Education: Artificial Intelligence.
The queries were executed on September 2023, including articles from 2010 to 2023, and they focus on the use of \textcolor{Mulberry}{sequential analysis} in an \textcolor{Mahogany}{educational setting}, where the obtained patterns are leveraged for an \textcolor{BurntOrange}{educational intervention}.
Two queries were executed, one per database, using the query below, with the results described in Section \ref{sec:dataprocess}.

\begin{center}
  \textit{ \textcolor{Cerulean}{("data mining" OR "artificial intelligence" OR "learning analytics" OR "intelligent tutor*" OR "learner model*")}  
  AND \textcolor{Mulberry}{ ( sequen*  OR  procedur*  OR  series )} 
  AND \textcolor{Mahogany}{( educatio*  OR  student  OR  teacher  OR  instructor )} 
  AND \textcolor{BurntOrange}{( feedback  OR  assessment  OR  predict* )}}
\end{center}

\subsection{Study Selection Process and Eligibility Criteria:}
The results were included if they were full-length articles, published in a peer-reviewed source, and written in English. 
This was followed by two rounds of screening, the first one based on the abstracts and the second on full text for those that were unclear from the first screening.
To be included, the publication had to be implemented in an educational setting, where the learning sequences refer to an ordered series of actions during a learning task.
Non-empirical works are excluded, as are works that only mention education as part of the institution name or a possible use case.
Additionally, publications were excluded if they analysed unrelated sequencing, like proteins, or where sequences were part of the methodology, such as “a sequence of trials” or “a series of interviews”.
Finally, works exploring learner patterns across multiple tasks were excluded, such as navigation patterns on the elements accessed in MOOC or the courses in a curriculum.

\subsection{Data Collection Process and Data Items}
\label{sec:dataprocess}
\begin{wrapfigure}{r}{0.45\textwidth}
    \includegraphics[width=\linewidth]{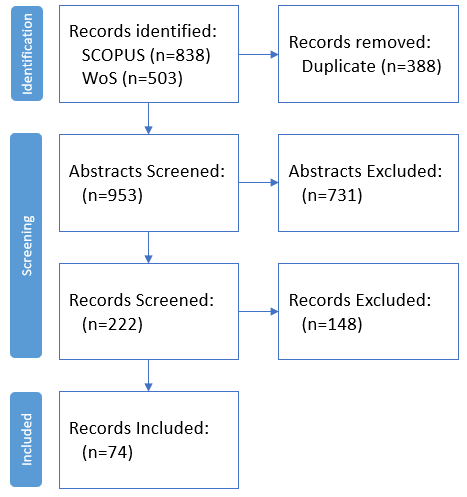}
    \caption{Document Flow Diagram}
    \Description[Document selection Flow Diagram]{Selection flowchart from 1341 documents to the final 74}
    \vspace{-0.9cm}
    \label{fig:prisma}
\end{wrapfigure}
Reference lists were exported from SCOPUS and WoS, and then transferred to Google Sheets for screening. 
The reason for exclusion or inclusion was recorded for each article, and the values are shown in Figure \ref{fig:prisma}. 
Of the articles obtained from the queries, 953 were unique, then in the abstract screening 717 were excluded since they were not in an educational setting, not empirical work, the use of learner action sequences was not part of the analysis or did not focus on a single learning task.
The full text of the remaining 236 articles was screened, and 162 were excluded based on the criteria described above, resulting in 74 articles included.
Document collection and inclusion were performed by the first author, using the criteria previously defined with the rest of the team. 

\subsection{Analysis Procedure} 
The included studies were coded by the first author using ATLAS.ti (a qualitative research software) with the Research Questions from Section \ref{sec:objectives}) as a guide, following negotiated consensual validation with the rest of the authors to iteratively review the codes \cite{sandelowski2007HandbookSynthesizingQualitative}.
For each question, we determined 3 or 4 possible categories from the related work in Section \ref{sec:related}, and then the full text of the articles was reviewed and coded using the questions and categories. 
If a code answered one of the questions but did not fit any of the categories, a new category was created.
Additionally, a second coder reviewed a sample of 10\% of the articles and the results were compared to ensure the reliability of the coding process, which showed an inter-rater reliability kappa of 0.922, where any differences were discussed and addressed.

For the first question, we identified the learning tasks considering the educational construct they are based on: deliberate practice, solving a problem or scientific inquiry \cite{wang2023SystematicReviewEmpirical}.
The first one refers to tasks where learners practice a skill through repetition, usually with a system tracking their progress and adjusting accordingly.
The second one refers to tasks where learners have to solve a complex problem in a virtual environment, and the third one refers to tasks where learners have to obtain and aggregate information.

On the other hand, the purpose of sequence analysis can be defined similarly to the goal of data mining and analytics tasks, such as clustering, classification or prediction \cite{xiao2022SurveyEducationalData}, in line with the findings in \cite{wang2023SystematicReviewEmpirical}.
While all the reviewed works identify patterns using sequence analysis, some of them go further and compare the patterns between groups of learners, for example, to identify behaviour differences among learners who received different experimental treatments, or to classify them by the use of strategies.
Finally, some works use the identified patterns to predict the next action of a sequence, the success of such action, or a learner's state. 

Regarding the units of the sequences, the reviewed research uses data sources such as log files or event records \cite{winne2010ImprovingMeasurementsSelfRegulated}, which have to be mapped to representations of meaningful actions, to be analysed and understood as part of the learning sequence \cite{nesbit2007AdvancingLogAnalysis}.
These sequence units can categorised as learning actions, which are domain or system-specific actions such as selecting and using a system tool in an OELE \cite{wang2023SystematicReviewEmpirical}, or metacognitive actions, usually defined based on an educational theory, such as self-explanation  \cite{kinnebrew2011ModelingMeasuringSelfregulated}.
The third category is problem attempts, the series of submissions in a guided practice system \cite{xiong2016GoingDeeperDeep, nadimpalli2023SystematicLiteratureReview}. 

For the implemented methods, we obtained the categories of methods that use sequential data from related literature on AI in Education and Process Mining \cite{nadimpalli2023SystematicLiteratureReview, bogarin2018SurveyEducationalProcess}: Probability Modeling, Deep Learning, Sequential Pattern Mining, Educational Process Mining and Knowledge Tracing.
Additionally, we identify the scope of the resulting sequences as the reference for interpretation, from system-specific tactics to general strategies for a complete task \cite{bogarin2018SurveyEducationalProcess}.

Regarding interventions and educational theories, we seek to identify which works go beyond the observation and exploration stage, which is why we included \textit{assessment, prediction and feedback} in our query, in Section \ref{sec:methods}.
However, considering the challenge of implementing and evaluating an educational intervention \cite{rienties2017ImplementingLearningAnalytics, motz2023LAKDirectionMisalignment}, we classify if articles implement an intervention, or if they explicitly mention a potential intervention or not.
Additionally, the impact of interventions relies on a strong theoretical foundation \cite{wise2015WhyTheoryMatters}, so we identify the educational theories used to support or interpret the results of sequential analysis \cite{khalil2022UseApplicationLearning}.

\subsection{Limitations}
The search was limited to Web of Science and SCOPUS, while these libraries include a wide range of publications, the results are not exhaustive.
Additionally, there are no established practices to mention sequence analysis in the title, keywords, or abstract, in consequence, some articles may not have appeared in our queries.
Furthermore, abstracts were used to assess the first round of inclusion, there is the possibility that articles were erroneously excluded.
Regarding the analysis, the articles were coded only by the first author, however, we addressed this limitation by following negotiated consensual validation with the author team \cite{sandelowski2007HandbookSynthesizingQualitative} and assessing the agreement of a sample with a second coder.


\section{Results}
\begin{figure}[ht]\centering
    \includegraphics[width=\linewidth]{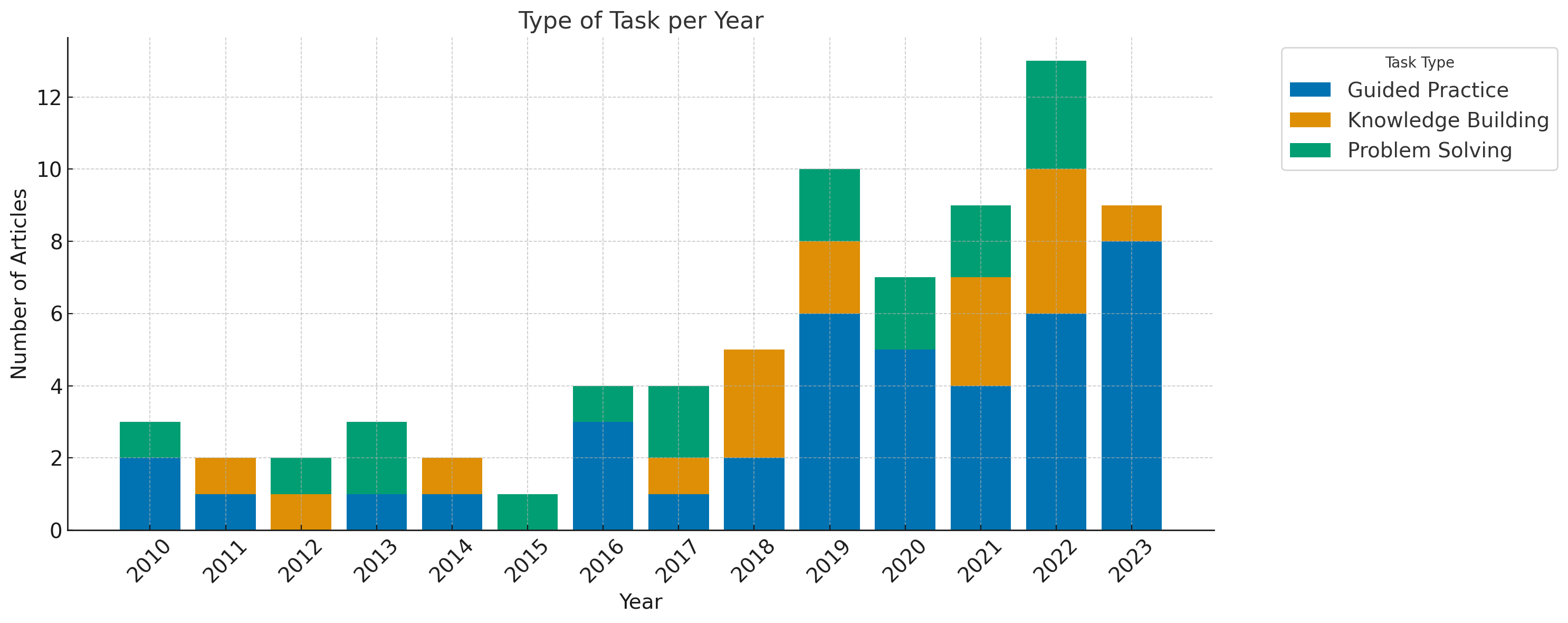}
    \caption{Articles, and Task Types, per Year}
    \Description[Bar chart of Articles, and Task Types, per Year]{Stacked, vertical, bar chart of the number of articles published per year, aggregated by task types, going from 2010 to 2023}
    \label{fig:taskyear}
    \vspace{-0.6cm}
\end{figure}

Our analysis included 74 articles, with the number of publications doubling in the second half of our search range, suggesting a constant increase in interest in the field, shown in Figure \ref{fig:taskyear}.
The articles were analysed following the Research Questions in Section \ref{sec:objectives}, to establish the trends of the different types of learning tasks, the different purposes of the sequence analysis, the types of units and analysis methods used, and any implemented interventions. 
The following sections address each of these questions, with the results available as an open-access dataset \cite{valletorre2023CodedDatasetLiteraturea}, and as an interactive interface at \href{https://sequence-dashboard.learn.ewi.tudelft.nl/}{https://sequence-dashboard.learn.ewi.tudelft.nl/}.

\subsection{Which types of learning tasks are analysed using sequences?}
\label{sec:tasks}
This review focuses on tasks that require active engagement from the learner, which instructional design models such as the 4C/ID model \cite{kirschner2008TenStepsComplex}, consider as authentic problems and practices.
However, they can follow different educational designs.
We identify three main types of tasks: guided practice, problem-solving, and knowledge building; the yearly number of publications is shown in Figure \ref{fig:taskyear}. 
The first type is \textit{guided practice}, characterized by a set of practice questions that are part of a topic, where a system uses the sequence of answers to constantly measure and adapt to the learner's knowledge level.
We registered 40 articles of this type, 2 of them using dialogue \cite{gonzalez-brenes2011HowClassifyTutorial, myers2021AutomaticDetectionStudent}, 
and 2 as a quiz \cite{juhanak2019UsingProcessMining, qu2019PredictingStudentAchievement}.

The second type of task is \textit{problem-solving}, a complex task integrating domain knowledge with domain-specific and general skills. 
In these tasks, the learner's interactions with the system can be interpreted as the operations executed towards a solution state \cite{anderson2000CognitivePsychologyIts}. 
While these tasks generally have a defined objective, they can be satisfied with different sequences of actions or applied operators, from a wide variety of actions available within the educational system.
The 17 articles using data from a variety of problem-solving tasks, including programming assignments \cite{kesselbacher2019DiscriminatingProgrammingStrategies}, mechanical engineering problems \cite{smith2015DiagrammaticStudentModels} or procedures in a biotechnology virtual lab \cite{riofrio-luzcando2016StudentActionPrediction, riofrio-luzcando2017PredictingStudentActions}.

Finally, some of the tasks require that the learner collects and processes information, from video lectures or reading materials, and then generates and evaluates a representation of knowledge, such as a concept map \cite{bouchet2012IdentifyingStudentsCharacteristic}.
In these tasks, the final result may be similar for learners who followed different strategies, or sequences of steps, and would benefit from personalized feedback. 
We identified 17 articles that can be classified as \textit{knowledge-building} tasks, for example, where a learner interacts with articles on climate change and then generates a causal model to explain the relationships between the learned concepts \cite{omae2018DataMiningDiscovering}. 

In summary, we identified 40 guided practice tasks, 17 problem-solving and 17 knowledge-building tasks, with the first type gaining ground in recent years, shown in Figure \ref{fig:taskyear}.
This may be due to the introduction of increasingly sophisticated deep learning methods or the availability of public datasets \cite{xiong2016GoingDeeperDeep, wu2023FusingHybridAttentive, ni2023HHSKTLearnerQuestion}, eliminating the need to design new systems or collect data.
Additionally, it might be related to the research objectives, described below.

\subsection{What is the motivation or \textit{purpose} of using learning sequences for analysis?}
\label{sec:purpose}
There are many affordances to the use of learning sequences and sequence analysis methods in EDM and LA, such as the ability to extract the sequential patterns from data \cite{lundgren2022LatentProgramModeling}, to identify them as tactics or strategies \cite{morshedfahid2021ProgressionTrajectoryBasedStudent}, or to include temporal structures in behaviour analysis \cite{li2022TemporalStructuresSequential}. 
Most articles identify the patterns as a learning behaviour, tactic or strategy, however, some of them go further and use these patterns for classification or for prediction. 
We found 8 cases of \textit{pattern identification}, for example, as the self-regulatory processes that occur before a help-seeking event.
These patterns can be used to anticipate the struggle of learners who may not use help-seeking tools correctly \cite{lajoie2015ModelingMetacognitiveActivities}. 
Another case of identification is when groups of learners receive different experimental treatments, such as a feedback strategy \cite{sun2019EffectsIntelligentFeedback}, and then the patterns of the groups are compared.
This means that in these 6 cases, the identified sequences are used as part of the evaluation of an educational intervention \cite{kinnebrew2014DifferentialApproachIdentifying}.

Going a step further, in the 24 articles where the purpose of the sequence is to serve for \textit{classification}, researchers first separate the learners by a given characteristic, such as proficiency in the task \cite{rajendran2018TemporalModelLearner} and then identify the patterns per group.
In these cases, like in traditional data mining classification tasks, a small percentage of the data is used to evaluate the accuracy of the method, or how well can it classify a learner by their behaviour \cite{rhienmora2010VirtualRealitySimulator, rajendran2018TemporalModelLearner, herold2013MiningMeaningfulPatterns}.
In only one article the grouping is done by applying clustering algorithms on the sequences of all learners, and then samples of the groups are reviewed and labelled as a certain strategy, such as \textit{Efficient Completion} \cite{morshedfahid2021ProgressionTrajectoryBasedStudent}.

Regarding \textit{prediction}, the purpose is to use a sequence as input but there is a variety of outputs: in 8 cases it is the learner's performance, such as success or failure on the task \cite{smith2015DiagrammaticStudentModels, chu2021ClickBasedStudentPerformance}, in 2 is the learner's state such as emotional or cognitive states \cite{corbeil2020ProbabilisticApproachesDetect, baker2010DetectingGamingSystem} and for 1 it is the next action they might do \cite{riofrio-luzcando2016StudentActionPrediction}.
For those works focusing on prediction, this is often with the vision of an early warning system or even automated interventions.
Finally, in 24 cases, the purpose is to predict the success of a learner's next action, often to determine their knowledge level for a given topic \cite{vie2019KnowledgeTracingMachines}.


\subsection{Which actions are used as the \textit{units} of learning sequences?}
In the analysed articles we found a broad variety of data records logged, including clicks on a web interface, sensor data from haptic devices, and text input in command line interfaces. 
A single case uses the data as-is, as the authors consider them detailed enough for the sensor data collected from a dentistry virtual reality simulator \cite{rhienmora2010VirtualRealitySimulator}.
In the rest of the publications, we found three different types of combining data records into traces: learning actions, problem attempts, and meta-cognitive actions.
We identified 26 articles that use \textit{learning actions} as units, which usually represent the different ways that learners apply operators in \textit{problem-solving tasks}, defined in Section \ref{sec:tasks}, such as editing code, executing a command, or using a tool.
Mapping the data records into such actions allows researchers, teachers and students to easily analyse and interpret the results of sequence methods.
The downside is that this mapping usually requires expertise with the domain and system, to know which actions are possible, and it also limits the possibility to generalize the analysis and results.

In the second case, identified in 37 of the reviewed works, the collected data records are better portrayed as \textit{problem attempts} because they are always delimited by a submission event: when a student submits a response to be evaluated by the system, usually named Intelligent Tutoring System (ITS), and they correspond to those using the \textit{guided task} type.
Considering that there is a clear definition of where one unit starts and ends, there is little focus on the process of mapping data records to units.

Finally, for the 10 articles using \textit{meta-cognitive actions}, we found that they introduced a conceptual background from different learning theories, and then use the selected theory to construct the sequence units. 
Eight of them use a Self-Regulated Learning (SRL) model to map certain records into a meta-cognitive action.
From the rest, one maps the data to meta-cognitive states, which they refer to as on-task and off-task states \cite{sun2019EffectsIntelligentFeedback}, and the other to a variety of meta-cognitive and cognitive behaviours, such as evaluating, help-seeking, and skipping video \cite{chen2022ExploringTimeRelatedMicroBehavioral}.
Using conceptually mapped actions reduces the granularity and length of the sequences to be analysed since they usually aggregate data records into just a few possible options, compared with the simpler learning actions above. 
Additionally, this means less variability during analysis and a more direct interpretation. 

The sequence units used for analysis are strongly related to the type of learning task, the 26 articles with learning actions tend to be on problem-solving, using specific actions in specific problems, and the 37 using problem attempts are generally on guided practice tasks, focusing more on the accuracy of the learner's model. 
Finally, the 10 cases of meta-cognitive actions are usually in the knowledge-building tasks, focusing more on self-regulated learning processes.

\subsection{Which \textit{methods} are used for the analysis of learning sequences?}
The methods in the reviewed articles usually take a set of sequences as input, where each one corresponds to learning actions executed by one learner during a task.
The methods that we found range from the manual analysis of game logs \cite{owen2016ModelingUserExploration} to deep learning models capable of predicting the performance of the learner \cite{smith2015DiagrammaticStudentModels}, with many others among them.
In our analysis, we classify the different methods into 5 categories: Sequential Probability Models, Educational Process Mining, Deep Learning, Knowledge Tracing, and Sequential Pattern Mining (SPM).
\begin{table}[ht]
    \centering
    \begin{tabular}{|p{2cm}|p{2cm}|p{5.8cm}|p{4cm}|}
        \hline 
        Scope&  Method Category&  Analysis Method& References\\ \hline 
        \multirow{3}{*}{\parbox{1.99cm}{Macro-level Patterns}} & Educational Process Mining & Fuzzy mining, PetriNet, Process Maps & \cite{rajendran2018TemporalModelLearner, neyem2017UnderstandingStudentInteractions, juhanak2019UsingProcessMining} \\ 
        \cline{2-4}
        & Other Methods & N-gram processing, Network Visualization & \cite{lundgren2022LatentProgramModeling, hurtig2022NetworkVisualizationAssessment, hurtig2022VisualizationStudentsSolutions} \\ 
        \cline{2-4}
        & Sequential Probability Model & Markov Chain Model, Markov Chain Process, Hidden Markov Model, Microsoft Sequence Clustering, Markov Logic Networks, Probabilistic Automata and Hidden Conditional Random Fields & \cite{dever2023ComplexSystemsApproach, rhienmora2010VirtualRealitySimulator, ha2011GoalRecognitionMarkov, meliana2018AdoptingGoodLearnersPaths, omae2018DataMiningDiscovering, gonzalez-brenes2011HowClassifyTutorial, kinnebrew2011ModelingMeasuringSelfregulated, tang2021ModelingLearningBehaviors, dever2022PedagogicalAgentSupport, riofrio-luzcando2017PredictingStudentActions, riofrio-luzcando2016StudentActionPrediction} \\ 
        \hline 
        \multirow{3}{*}{\parbox{1.99cm}{Meso-level Patterns}} & Deep Learning & GRU, MLP, CNN, LSTM, SAE & \cite{chu2021ClickBasedStudentPerformance, smith2015DiagrammaticStudentModels, elaouifi2021PredictingLearnerPerformance, qu2019PredictingStudentAchievement, mubarak2021PredictiveLearningAnalytics, geden2020PredictiveStudentModeling, corbeil2020ProbabilisticApproachesDetect, zhou2022WhatStudentsInteractions} \\ 
        \cline{2-4} 
        & Knowledge Tracing & PFA, ModPFA, BKT, DKT, DKVMN, AKT, and many more & \cite{banno2020AutomatedCognitiveAnalyses, tuti2019BuildingLearnerModel, gong2010ComparingKnowledgeTracing, guo2023DeepKnowledgeTracking, zhang2023DynamicMultiobjectiveSequencewise, liu2023EnhancingDeepKnowledge, mandlazi2021EvaluatingDeepSequential, wu2023FusingHybridAttentive, xiong2016GoingDeeperDeep, ni2023HHSKTLearnerQuestion, lu2023InterpretingDeepLearning, he2020KTXLKnowledgeTracing, vie2019KnowledgeTracingMachines, jiang2022MAKTMultichannelAttention, gan2020ModelingLearnerDynamic, xia2023MultiVariateKnowledgeTracking, eagle2018PredictingIndividualizedLearner, sahebi2014PredictingStudentPerformance, schatten2016StudentProgressModeling, sahebi2016TensorFactorizationStudent, geng2023NoiseFilteringEnhancedDeep, yang2022SkillOrientedHierarchicalStructure, cook2017TaskTimingSeparating}\\ 
        \cline{2-4} 
        & Other Methods & Dynamic Time Warping, Model-free bounded optimization, Course-agnostic Student Performance Model & \cite{morshedfahid2021ProgressionTrajectoryBasedStudent, chen2018ModelFreeEstimateLimits, schmucker2022TransferableStudentPerformance}\\ 
        \hline
        \multirow{3}{*}{\parbox{1.99cm}{Micro-level Patterns}} & Sequential Pattern Mining & SPAM, Differential, Temporal Interestingness of Patterns in Sequences (D-TIPS), Pex-SPAM, Generalized Sequential Pattern Mining, Frequent Item set,   & \cite{kinnebrew2014DifferentialApproachIdentifying, gomez2021ApplyingLearningAnalytics, myers2021AutomaticDetectionStudent, akhuseyinoglu2021DataDrivenModelingLearners, wang2019EvaluatingStudentLearning, gomez2020ExploringAffordancesSequence, taub2019HowDoesPrior, normlien2020HowWellTeachers, bouchet2012IdentifyingStudentsCharacteristic, linj.2022ItGoodMove, herold2013MiningMeaningfulPatterns, venant2017UsingSequentialPattern, latypova2022WorkFreeResponse, akhuseyinoglu2022ExploringBehavioralPatterns} \\ 
        \cline{2-4} 
        & Manual Analysis & Manual Labeling, Sequence of Action Tables, State Permutations  & \cite{duong2013PredictionModelUses, baker2010DetectingGamingSystem, kesselbacher2019DiscriminatingProgrammingStrategies, sun2019EffectsIntelligentFeedback}\\ 
        \cline{2-4} 
        & Other Methods & Visualization, Lag Sequential Analysis, GSEQ (Generalized Sequential Querier), Time-embedded n-gram & \cite{supianto2019GTRASGraphicalTracking, liu2018ApplyingLearningAnalytics, li2022TemporalStructuresSequential, chen2022ExploringTimeRelatedMicroBehavioral}\\ 
        \hline
    \end{tabular}
    \caption{Methods per Category and Scope}
    \label{tab:methods}
    \vspace{-0.8cm}
\end{table}
\textit{Sequential Probability Models} estimate the probabilities of transition between elements in a sequence, based on the previous one or more elements \cite{dever2023ComplexSystemsApproach}.
On the other hand, \textit{Educational Process Mining} refers to the EDM technique to discover, visualise and compare process models, focusing on the complete sequence \cite{rajendran2018TemporalModelLearner}. 
\textit{Deep Learning} methods use Artificial Neural Networks (ANN) to predict the outcome of a sequence, mainly Recurrent Neural Networks (RNN) and Long Short-Term Memory (LSTM) networks, as they are better suited to handle sequential data than other ANNs \cite{chu2021ClickBasedStudentPerformance}.
While \textit{Knowledge Tracing} approaches leverage probabilistic, machine learning or deep learning techniques, their main focus is to model the knowledge evolution of a learner throughout their interaction with an educational system \cite{guo2023DeepKnowledgeTracking}.
\textit{Sequential Pattern Mining} methods refer to a group of data mining techniques used to find statistically significant patterns---or subsequences---in sequence datasets, such as actions that happen frequently in the same order across learners' records \cite{kinnebrew2014DifferentialApproachIdentifying}.
\textit{Manual Analysis} methods are often used in combination with other methods but are also used on their own in a few cases, for instance by reviewing and classifying the actions of individual learners in a small group \cite{baker2010DetectingGamingSystem}.
These methods often relate to the scope of the analysis, from the full-task strategies shown by a learner population to short tactics that certain students apply depending on context: the Macro-level Patterns identify strategies that span across the complete task and are analysed for a population or group.
The Meso-level refers to behaviours that appear in large part of the task, but the focus is on the individual level.
Finally, the Micro-level patterns are short tactics that occur in a particular context.
Shown in Table \ref{tab:methods}, we group the examples of each method and the scope that they belong to.

\subsection{Educational Interventions - What is the final intervention designed using the obtained insights?}
\label{sec:intervention}
Educational interventions provide students with the support required to acquire the skills and knowledge being taught.
However, interventions remain one of the main challenges in the learning analytics cycle \cite{rienties2017ImplementingLearningAnalytics}. 
This is reflected in our findings, where most reviewed publications are exploratory studies with potential interventions, for instance, as an early warning system that can notify teachers about learners using tactics that lead to poor performance \cite{wang2019EvaluatingStudentLearning}, or automated systems that can detect if a learner might go into a blocking state \cite{corbeil2020ProbabilisticApproachesDetect}.
There were also 5 prototype teacher dashboards \cite{juhanak2019UsingProcessMining, gomez2020ExploringAffordancesSequence, gomez2021ApplyingLearningAnalytics, supianto2019GTRASGraphicalTracking, bryfczynski2013TeachingDataStructures}, with one of them receiving a positive response from a usability survey with 56 participants \cite{hurtig2022VisualizationStudentsSolutions}.

In our review, we found 2 articles that design and implement an intervention. 
In the first, the authors integrate the patterns obtained to provide suggestions to students in a programming tutoring system.
The suggestions included reading the manual or asking a peer for help when an error state is identified, to nudge the learners into a self-regulation strategy, for example, \textit{reflection-then-success} after a sequence that could lead to failure \cite{venant2017UsingSequentialPattern}.
The impact of this intervention, however, is not evaluated.
The second one implemented and evaluated the use of learning sequences, by creating a ``good-learner path" model using a Markov Decision Process in a guided practice system, so it can automatically generate the content sequence for first-time users. 
The model was then evaluated by comparing the test results of learners who followed the created path and learners who were provided with exercises one by one based on their prior results.
Their approach produced 19.5\% more good learners and 20.5\% lower standard deviation than the control group \cite{meliana2018AdoptingGoodLearnersPaths}.

\section{Discussion}
\subsection{Learning Tasks and Analysis Purpose}
Regarding the \textit{purpose} of sequence analysis, the common objective is to extract patterns from the data and identify the real-world behaviours they describe. 
In some cases, these patterns are then used to classify the learner between groups or to predict the learner's state, their subsequent action, or the outcome of the task.
In most research using problem-solving tasks, patterns are closely related to the scenario, for example, how novices implement functions in programming.
A few cases use these scenario-specific patterns to understand more general skills, for instance, identifying if learners follow linear or iterative processes for problem-solving.
Similarly, for the knowledge-building tasks, the focus is to find information processing strategies and the meta-cognitive processes executed by the learners \cite{kinnebrew2011ModelingMeasuringSelfregulated, kinnebrew2014DifferentialApproachIdentifying}.
Using an educational lens to determine the purpose of a study, such as complex problem-solving or self-regulation, facilitates the generalisation of the methods and results, regardless of the task.

On the other hand, the purpose of many guided practice tasks is to improve the prediction of the learners' performance, often to evaluate a proposed method against the state-of-the-art, as is the case for most of the 16 articles using the ASSISTments\footnote{https://sites.google.com/site/assistmentsdata/datasets} dataset. 
In these studies, the sequential patterns receive less attention, and educational theory is leveraged in the prediction function, such as learning and forgetting, or while calculating the difficulty of a learning component.
While this field has seen important growth with the popularity of Neural Networks (see figure \ref{fig:taskyear}), the real-world application and explainability of modern techniques are still key challenges \cite{song2022SurveyDeepLearning}.

\subsection{Learning Actions and Methods}
From our analysis, there seem to be increasing similarities in how researchers of different domains use sequences for the analysis of learner behaviour.
For example, we found that 25 out of the 37 articles using Problem Attempts as the units of sequence, also integrate contextual information about the learner and their actions before submitting a problem, such as temporal features or their previous knowledge.
Another common case is the use of statistical methods in combination with sequential methods, such as using logistic regression with features extracted from a sequential technique.
Furthermore, researchers often implement more than one sequence analysis technique, using one to find case-specific action units, and then another to explore how these actions are sequenced in a full task strategy \cite{tang2021ModelingLearningBehaviors}.
This may be a consequence of recent efforts towards the transparency and explainability of data-driven analysis, as well as the increasing accessibility of technology and analysis techniques.

The use of machine learning and deep learning methods to power automated or semi-automated decisions has recently started to see widespread implementation. 
This may be due to the recent efforts to mitigate risks associated with AI, for instance by pursuing transparency, causality, privacy, fairness, trust, usability, and reliability; some of the goals of Explainable AI (XAI) \cite{lu2023InterpretingDeepLearning}.
Explainability is especially important when applied to education, as the explanations can also have an educational component for learners, and it can aid teachers in their assessment of learners and their courses \cite{lu2023InterpretingDeepLearning}. 
Regarding accessibility, different techniques are available via tools and libraries, we found several articles that use the Waikato Environment for Knowledge Analysis (Weka\footnote{https://www.cs.waikato.ac.nz/ml/weka/}) for multiple machine learning classifiers and predictors, \cite{herold2013MiningMeaningfulPatterns, myers2021AutomaticDetectionStudent, sahebi2014PredictingStudentPerformance, riofrio-luzcando2017PredictingStudentActions, owen2016ModelingUserExploration, padron-rivera2015IdentifyingAffectiveTrajectories, bouchet2012IdentifyingStudentsCharacteristic}, 2 more that apply Sequential Pattern Mining with TraMineR for R\footnote{http://traminer.unige.ch/} \cite{linj.2022ItGoodMove, tuti2019BuildingLearnerModel}, as well as Markov Models via MeTA Toolkit\footnote{https://github.com/meta-toolkit/meta} \cite{tang2021ExploratoryAnalysisLatent}, and the Microsoft Sequence Clustering Algorithm \cite{riofrio-luzcando2017PredictingStudentActions}.
It is important to note that these tools were explicitly mentioned in their respective articles, further facilitating reproducibility and comparison.
This allows researchers to work with known methods throughout their analysis process, without having to develop, test, and evaluate them from scratch.
Finally, we provide the dataset \cite{valletorre2023CodedDatasetLiteraturea} of the reviewed works as an interactive interface \footnote{https://sequence-dashboard.learn.ewi.tudelft.nl/} where researchers can filter the methods depending on the scope of their study, as shown in Table \ref{tab:methods}.

\subsection{Interventions}
Some of the reviewed works mention potential interventions that can be designed using the obtained patterns.
For example, adaptive scaffolding in problem-solving tasks, personalized feedback for a knowledge-building system, or detection of the learners' level in guided practice.
However, only 2 implemented an intervention, with 1 of them evaluating it, while 20 articles had no mention of a potential intervention at all.
This is in line with recent publications such as "LAK of Direction", where they found that 11\% of their reviewed articles introduced an intervention into a learning environment \cite{motz2023LAKDirectionMisalignment}.  

The use of learning principles, such as cognitive load \cite{myers2021AutomaticDetectionStudent} or learning and forgetting curves \cite{gan2020ModelingLearnerDynamic}, is often present when describing or framing the problem.
Very few use learning principles in the interpretation of patterns or interventions.
This may be due to the focus on task- or system-specific tactics more than general learning behaviours.
However, actionable results in LA research require a strong theoretical foundation, present in the analysis and the interpretation of the findings \cite{wise2015WhyTheoryMatters}.
The theories mentioned in the framing of the problem should remain through the complete process, such as the use of Neo-Piagetian theory to classify patterns as programming strategies related to expertise \cite{kesselbacher2019DiscriminatingProgrammingStrategies}.
This was the case for the 10 articles where the purpose is to identify meta-cognitive strategies, with 8 of them using a Self-Regulated Learning model to map the data records into activities, interpret the patterns after analysis, and describe a potential intervention \cite{khalil2022UseApplicationLearning}.

\section{Conclusion}
In this document, we reviewed the different learning tasks, research purposes, action units, and methods of learning sequence analysis in educational systems, as well as the interventions that researchers design and propose with the obtained insights. 
We described the major trends that can help to establish an overview of the use of Learning Sequence Analysis, aggregated in an open repository \cite{valletorre2023CodedDatasetLiteraturea} and an interactive interface available in \href{https://sequence-dashboard.learn.ewi.tudelft.nl/}{https://sequence-dashboard.learn.ewi.tudelft.nl/}.
This is to assist researchers in finding, comparing, and using other's efforts as a foundation and move past the exploratory stage; as noted in one of the reviewed articles \cite{riofrio-luzcando2016StudentActionPrediction}, "the community is missing more generally applicable results."

\begin{acks}
  This project has been funded by the Leiden-Delft-Erasmus Centre for Education and Learning (LDE-CEL). The authors thank the second coder, Rebecca Glans.
\end{acks}

\bibliographystyle{ACM-Reference-Format}
\bibliography{references}

\end{document}